\documentstyle[twocolumn,prb,aps,grafik]{revtex}

\begin{document}
\draft
\twocolumn[\hsize\textwidth\columnwidth\hsize\csname@twocolumnfalse\endcsname
\vspace{0.5cm}
\begin{flushright}
Submitted to {\bf PHYSICAL REVIEW B}
\end{flushright}
\title{Energy-transfer rate in a double-quantum-well system 
due to Coulomb coupling}
\author{R.\,T. Senger and B. Tanatar}
\address{Department of Physics, Bilkent University, Bilkent, 06533 
Ankara, Turkey}
\maketitle

\begin{abstract}
We study the energy-transfer rate for electrons in a 
double-quantum-well structure, where the layers are coupled through 
screened Coulomb interactions. The energy-transfer rate between the 
layers (similar to the Coulomb drag effect in which the momentum 
transfer rate is considered) is calculated as functions of electron 
densities, interlayer spacing, the temperature difference of the 
2DEGs, and the electron drift velocity in the drive layer. We employ 
the full wave vector and frequency dependent random-phase 
approximation at finite temperature to describe the effective 
interlayer Coulomb interaction. We find that the collective modes 
(plasmons) of the system play a dominant role in the energy transfer
rates. The contribution of optical phonons to the transfer rates 
through the phonon mediated Coulomb coupling mechanism has also 
been considered.
\end{abstract}
\vskip1pc]

\narrowtext
\newpage

\section{Introduction}

Coupled quantum-well systems are known to exhibit rich and 
interesting physics, where correlation effects are 
significant.\cite{hanna} In particular, the Coulomb drag effect 
is a unique way of probing many-body correlations through a 
transport measurement,\cite{rojo,flensberg95,hill97} where one of 
the layers is driven by an external current, and the influences on
the other (drag) layer are investigated. The interlayer
carrier-carrier interactions lead to measurable effects, such as
transresistivity due to momentum transfer between the layers.
The observed transresistance crucially depends on the single-particle
and collective excitations of the coupled system, because these
excitations are the ones which mediate the momentum and energy
transfer between the layers. There has been a growing 
theoretical\cite{zmac,jauho93,tso,kamenev}
and experimental\cite{tulipe,gramila,sivan,noh} activity in the 
past years touching upon various aspects of the drag phenomenon.

In this paper we study the energy transfer between two layers of
quasi-two-dimensional electron gases under experimental 
conditions similar to the transresistivity measurements. 
The importance of the energy transfer between two Coulomb coupled 
quantum wells were pointed out by Price.\cite{price83,price88}
In the hot-electron context, the energy transfer occurs when
there is a difference of electron temperatures in the two
layers. In the actual drag experiments\cite{tulipe,solo91} the
energy transfer was detected from the heating effects. 
The energy transfer rate in spatially separated systems were
theoretically considered also by Jacobini and
Price,\cite{jacobini} Laikhtman and Solomon,\cite{laik} 
Boiko and Sirenko,\cite{boiko} and recently by
Tanatar\cite{tanatar97} who considered the case of a coupled
quantum wire system.

We calculate the temperature dependence of the energy transfer rate 
in a double quantum well system. It is assumed that the wells may 
be kept at different carrier temperatures which are also different 
from the lattice temperature in general.\cite{cui93} The 
calculations are based on the random-phase approximation
(RPA), with full consideration of wave vector and
frequency dependencies at finite temperatures. The layers are
coupled through Coulomb interactions, and in the steady state the 
resulting charge polarization produces an electrostatic field which 
compensates the drag force in the drag layer. Using the
momentum and energy balance equations we investigate 
the static and dynamic screening effects on the power transfer
between the layers.
In the drive layer the influence of the externally applied electric 
field is treated in terms of the electron drift velocity. We
probe the effects of a finite drift velocity to study the
nonlinear regime of the energy transfer rate. The
nonequilibrium aspects of frictional drag has recently been
considered by Wang and da Cunha Lima,\cite{wang} who employed
the balance equation approach. The amount of transfered energy has a
direct dependence on the electron layer densities which can be 
different in general. We also calculate the effect of density 
mismatch on the energy transfer rates.
As a final point to be discussed in this report we consider the 
contribution of optical phonon exchange as an additional mechanism 
of interlayer interaction.   

\section{Model}

We consider two quantum wells of width $w$, and center-to-center
separation of $d$. The potential barriers are assumed to be
infinite, so that there is no tunneling between the layers. 
The two dimensional electron 
charge density in the first layer, $n_{1}$, is related to the Fermi
wave vector by $n_{1}=k_{F}^{2}/(2\pi)$, and $T_{F}$ is the 
corresponding Fermi temperature of the electron gas in the layer. 
It is also appropriate to define the dimensionless
electron gas parameter $r_{s}=\sqrt{2}/(k_{F}a_{B}^{\star})$, where
$a_{B}^{\star}=\epsilon_{0}/(e^{2}m^{\star})$ is the effective Bohr
radius in the layer material with background dielectric constant
$\epsilon_{0}$ and electron effective mass $m^{\star}$. 
For GaAs quantum-wells $a_{B}^{\star}\approx 100$\,\AA\ and
experimentally realized electron densities are of the order of
$10^{11}$\,cm$^{-2}$ which corresponds to $r_{s}\approx 1-2$, 
and $T_{F}\approx 40-100$\,K. We take the
charge density in the second layer with reference to the drive 
layer; $n_{2}=\alpha n_{1}$, so that the quantities for the second 
layer scale as, ~~$k_{F}^{(2)}=\sqrt{\alpha}k_{F}$, 
~~$r_{s}^{(2)}=r_{s}/\sqrt{\alpha}$,
~~$T_{F}^{(2)}=\alpha T_{F}$. 

The transport properties of the double
quantum well system can be characterized by the electron drift 
velocities $v_{i}$ and electron gas temperatures $T_{i}$. 
One of the layers (drive layer) is subject to an electric field in 
the $x$-direction which drives the electrons with a drift velocity 
$v_{1}$. The other well is kept as an ``open circuit'', therefore 
$v_{2}=0$. The drag experiments are performed at low electric
fields in the linear regime, so we shall take the limit
$v_1\rightarrow 0$ at the end of the calculations.
In this work our starting point for the calculation of the
energy transfer rate is the balance-equation approach to hot
carrier transport which has been successfully applied to a
variety of situations involving transport phenomena in
semiconductors.\cite{ting92} The resulting momentum and energy
transfer rate equations have also been obtained within a variety
of other techniques.\cite{flensberg95,zmac,jauho93,boiko}
With the assumption that only the lowest subband in each layer is 
occupied, the momentum and energy transfer rate expressions due
to interlayer Coulomb interactions,
derived within the balance equation approach to nonlinear 
electrical transport in low dimensional semiconductors,
are given by\cite{cui93,wang,ting92} ($\hbar=k_{B}=1$);
\begin{eqnarray} 
f_{12}(v_1-v_2) &=&
 -\sum_{\vec{q}} q_{x}\,\int_{-\infty}^\infty
{d\omega\over\pi}\,\mid W_{12}(q,\omega)\mid^2 \nonumber \\
 &\times~& 
\left[ n_{B}\left(\frac{\omega}{T_1}\right)
      -n_{B}\left(\frac{\omega-\omega_{12}}{T_2}\right) \right] 
   \nonumber \\
 &\times&~ {\rm Im} \chi_{1}(q,\omega)~
         {\rm Im} \chi_{2}(-q,\omega_{12}-\omega)~,
\end{eqnarray}
and
\begin{eqnarray}
P_{12} (v_1-v_2) &=&
 -\sum_{\vec{q}} \int_{-\infty}^{\infty}\frac{d\omega}{\pi}~\omega~
 \mid W_{12}(q,\omega)\mid^2  \nonumber \\
 &\times&\left[ n_{B}\left(\frac{\omega}{T_1}\right)
      -n_{B}\left( \frac{ \omega- \omega_{12}}{T_{2}} \right) \right]
    \nonumber \\
 &\times&~ {\rm Im} \chi_{1}(q,\omega)~
         {\rm Im} \chi_{2}(-q,\omega_{12}-\omega)~
\end{eqnarray}
respectively.
In the above, $\omega_{12}=q_{x}(v_{1}-v_{2})$, 
$W_{12}(q,\omega)=V_{12}(q)/\varepsilon(q,\omega)$ is
the dynamically screened interlayer potential, Im$\chi(q,\omega)$ 
is the imaginary part of the temperature dependent 2D
susceptibility\cite{flensberg95} for a single layer, and 
$n_{B}(x)=1/(\exp(x)-1)$ is the
Bose distribution function. The screening function 
$\varepsilon(q,\omega)$ for the double-well system is written as
\begin{eqnarray}
 \varepsilon(q,\omega)&=&[1-V_{11}(q)\chi_{1}(q,\omega;T_1)]
[1-V_{22}(q)\chi_{2}(q,\omega;T_2)] \nonumber \\
   &-&V_{12}^2(q)\chi_{1}(q,\omega;T_1)\chi_{2}(q,\omega;T_2)\, ,
\end{eqnarray}
where
\begin{equation}
V_{ij}(q)=F_{ij}(qw)~\frac{2\pi e^2}{\epsilon_{0}q}~e^{-qd(1-
\delta_{ij})}
\end{equation}
define the intra- and interlayer unscreened Coulomb interactions, 
and $F_{ij}(qw)$ are the form factors\cite{flensberg95,jauho93}
for a model of infinite barrier and square wells of width $w$. 
Note that we have indicated
explicitly in Eqs.\,(1) and (2) that the layers or quantum-wells 
are kept at different temperatures. Under drag conditions
mentioned above, the interlayer resistivity
(transresistivity) reads
\begin{equation}
\rho_{12}=-{1\over n_1 n_2 e^2 v_1}\,f_{12}(v_1)\, ,
\end{equation}
where $f_{12}$ is the interlayer momentum transfer rate or
frictional force. Resistivity expression is further simplified 
if we consider layer temperatures to be
equal, $T_1=T_2$, and within the linear regime $v_1\rightarrow 0$,
yielding
\begin{equation}
\rho_{12}=-{1\over n_1 n_2 e^2}\,\left.{df_{12}(v_1)\over
dv_1}\right|_{v_1=0}\, .
\end{equation}
The energy transfer rate expression
given in Eq.\,(2) resembles the momentum transfer rate
expression of Eq.(1), except the transferred energy $\omega$ 
appears in the integrand, and the difference between the Bose 
distribution functions at different temperatures reduce to the 
familiar\cite{rojo,zmac,jauho93} $\sim 1/\sinh^{2}(\omega/2T)$ when 
$T_1$ approaches $T_2$. With the sign chosen in Eq.\,(2), $P_{12}$ is
the amount of power transferred 
{\it to} the layer 1 {\it from} the layer 2.

\section{Results and Discussion}

We first evaluate the energy-transfer rate $P_{12}(0)$ in the linear 
regime $(v_{1}=0)$ for a GaAs system. Even in this case, the
energy transfer rate is non-zero as long as the electron
gases are kept at different temperatures. 
The comparison of taking the 
interlayer potential as either statically or dynamically screened 
is presented in Fig.\,1(a). Because the energy transfer rate
$P_{12}$ changes sign as $T_2$ is scanned for a fixed $T_1$,
we plot $|P_{12}|$ in our presentations. In the statically screened
interaction we use $\varepsilon(q)=[1-V_{11}\chi_1(q)][1-V_{22}
\chi_2(q)]-V_{12}^2\chi_1(q)\chi_2(q)$, in which the static
response functions $\chi_{1,2}(q,\omega=0)$ enter.
Both quantum wells are taken of width $w=2a_{B}^{\star}$, and 
with equal electron densities $(n_{1}=n_{2})$.
The temperature of the first layer is kept at
$T_{1}=T_{F}$. For three distinct sets of $r_{s}$ and $d$ values, the
$T_{2}$ dependence of $|P_{12}|$ is plotted. We observe that the 
inclusion of dynamical screening effects yields qualitatively and 
quantitatively different results for the energy transfer rate. 
At very low $T_{2}$ both types of screening yield close results, 
but with the increasing temperature, $P_{12}$ significantly grows 
for dynamical screening, whereas it monotonically decreases for 
the case of static screening. The difference between the two 
approaches is attributed to the contribution
of plasmons which is completely missed for the statically screened 
interaction. Similar differences between the static and dynamic
screening approaches were also found in the momentum transfer
rate at high temperatures determining the 
transresistivity.\cite{flensberg95} We notice that the qualitative 
forms of the curves are roughly independent of the separation 
distance and the charge density. When the temperatures of both 
layers are equal the energy transfer vanishes for all cases in 
the linear regime.
This is also seen in Fig.\,1(b) where we take $T_{1}$ at three 
different values, and vary $T_{2}$ from 0 to $T_{F}$. 
Although experimentally the energy transfer is expected to be from
the drift layer to the drag layer (in the linear regime at least), 
for the sake of generality the
cases where $T_2 > T_1$ are also included in our plots. 
Note that $P_{12}$ changes sign when $T_{2}>T_{1}$, so that energy is
transferred from the hot layer to the cold one. When we compare the
overall behavior of $P_{12}$ in a double-layer system with that
in a coupled quantum wire system, we observe that in the latter
a pronounced peak structure\cite{tanatar97} around $T\sim 0.3\,T_F$ 
is present. On the other hand, the results shown in Figs.\,1(a) and 
1(b) indicate a rather broad enhancement coming from the plasmon
excitations. Price has estimated\cite{price88} the energy transfer 
rate (per electron) between coupled quantum wells to be $\sim
0.1-1$\,erg/s, for typical layer densities of $n\sim
10^{11}$\,cm$^{-2}$ and $d\approx 100$\,\AA. In obtaining this
estimate, the electron temperature was taken as $10^3$\,K, which 
is about $10\,T_F$. Our calculations are 
mostly done for layers of finite thickness, larger separation 
distances and at relatively lower temperatures around $T_{F}$. The 
results indicate rates of the order of
$10^{-2}$\,erg/s, which are much smaller than the estimate
given by Price.\cite{price88}

\bfig{h}\ff{0.45}{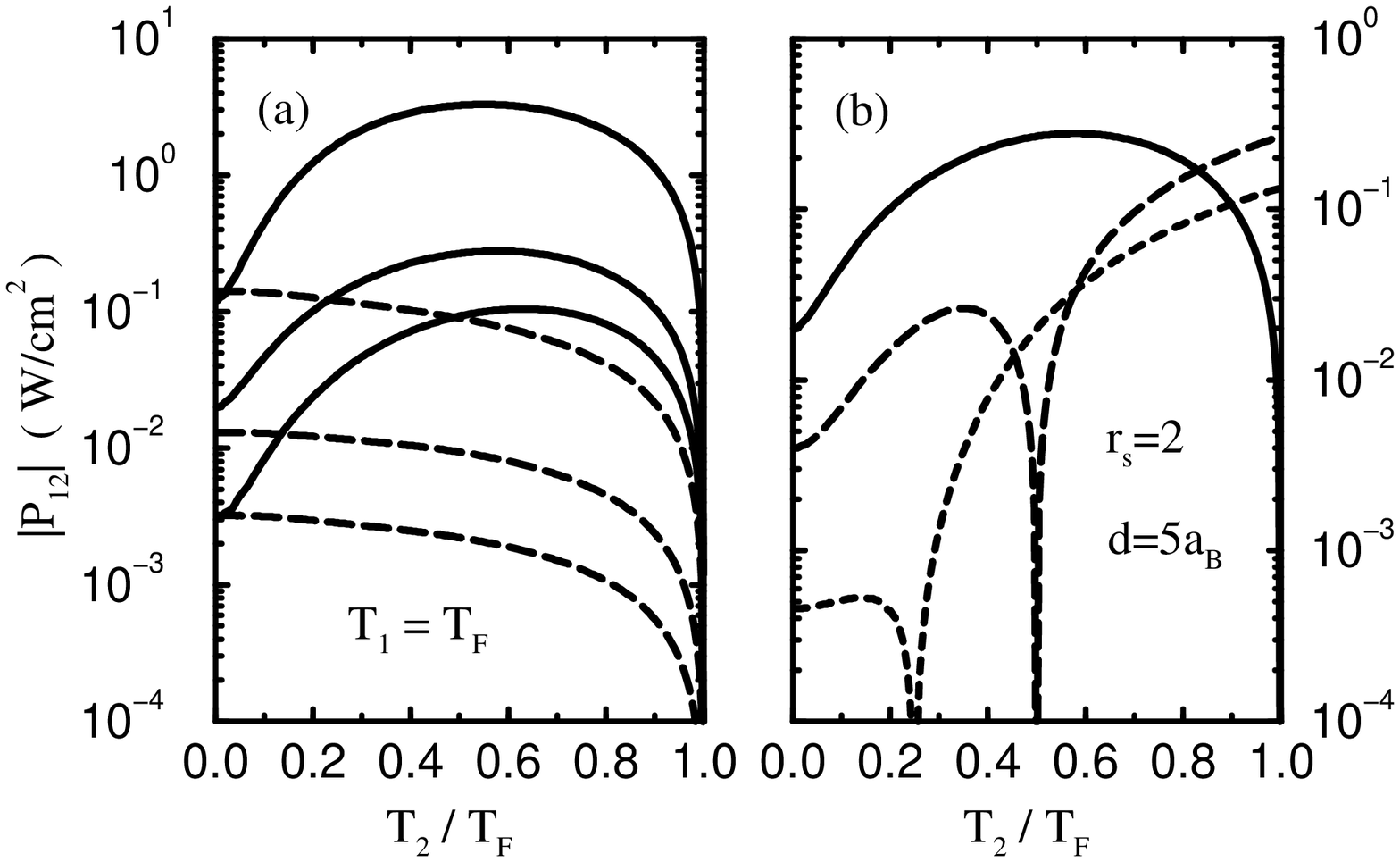}\efig{
The energy-transfer rate for identical quantum-wells of
width $w=2a_{B}^{\star}$, (a) in the static (dashed) and the dynamic
(solid) screening approximations as functions of temperature $T_{2}$.
The temperature of the drive layer is kept constant at $T_{1}=T_{F}$.
The couple of curves from top to bottom are for $(r_{s},d)$ values of
$(1,5a_{B}^{\star})$, $(2,5a_{B}^{\star})$ and $(2,7a_{B}^{\star})$
respectively, (b) for different values of $T_{1}$. The solid,
dashed, and short-dashed curves are for $T_{1}/T_{F}=1,\,0.5$, and
$0.25$, respectively.
}{f1}

\bfig{h}\ff{0.45}{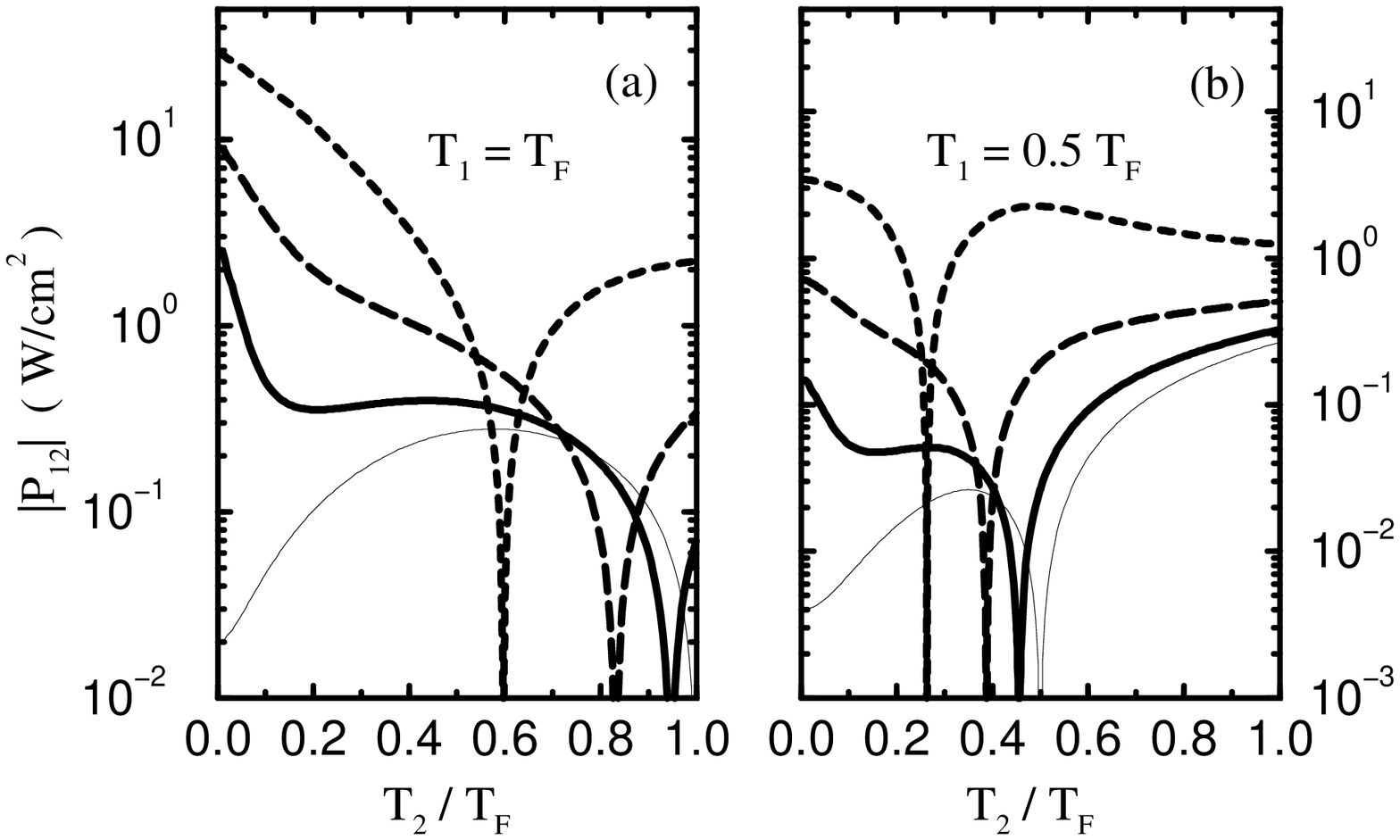}\efig{
The energy-transfer rate for identical wells
in the nonlinear regime when
(a) $T_{1}=T_{F}$, (b) $T_{1}=0.5T_{F}$. The solid, dashed, and
short-dashed curves are for $v_{1}k_{F}/E_{F}=0.5, 1$, and $2$,
respectively, in a $w=2a_{B}^{\star}$ and $d=5a_{B}^{\star}$
double-well system ($r_{s}=2$). The thin solid curves correspond
to the linear regime ($v_{1}=0$), included for comparison.
}{f2}

Next we investigate the energy-transfer rate in the nonlinear 
regime (i.e. for non-zero $v_{1}$). The density response function 
and the Bose distribution function of the drag layer are calculated 
at shifted frequencies $\omega-q_{x}v_{1}$. In Fig.\,2 the energy 
transfer rate is displayed in this nonlinear situation when $T_1$ is
kept at either $T_F$ or $0.5\,T_F$.
We observe that for a finite drift velocity and at very low 
temperatures of the second layer, the 
amount of transfered power is a few orders of magnitude larger than
that for the linear regime. As $T_2$ is increased, $|P_{12}|$ starts 
to decrease rapidly and eventually vanish at a critical value,
before the temperatures of the layers become equal. Larger the drift 
velocity lower the $T_2$ at which no energy is transfered between 
the layers (eg. for $T_1=T_F$ and $v_1=2E_F/k_F$, $P_{12}=0$ at 
$T_2\approx 0.6T_F$). Beyond that point, as a consequence of 
nonlinearity, for further larger values of $T_2$ the drive layer 
start to {\it absorb} power from the second layer even when 
$T_1>T_2$. 
The momentum transfer rate between two
coupled electron-hole quantum wells in the nonlinear regime was 
considered by Cui, Lei, and Horing.\cite{cui93} They found that
the nonlinear effects start to become significant at different
electric field strengths (or equivalently the drift velocity $v_1$) 
for different temperatures.
In a system of two sets of charged particles streaming relative to
one another the collective modes may undergo instabilities with
respect to charge density perturbations as studied by Hu and
Wilkins.\cite{huwilkins} It is conceivable that under the drag
effect conditions, such two-stream instabilities may be detected
for large drift velocities. Hu and Flensberg\cite{hcis}
predicted a significant rise in the drag rate just under the
instability threshold. Recent drag rate calculations of Wang and
da Cunha Lima\cite{wang} did not explore this phenomenon. In our
calculations of the energy transfer rate, 
the shift of vanishing $P_{12}$ point to lower $T_2$ values
is expected to reflect the onset of plasma instabilities. 
In our numerical results shown in Figs.\,2(a) and 2(b) it appears 
that two-stream instability limit is not yet reached for the drift
velocities chosen.

\bfig{h}\ff{0.45}{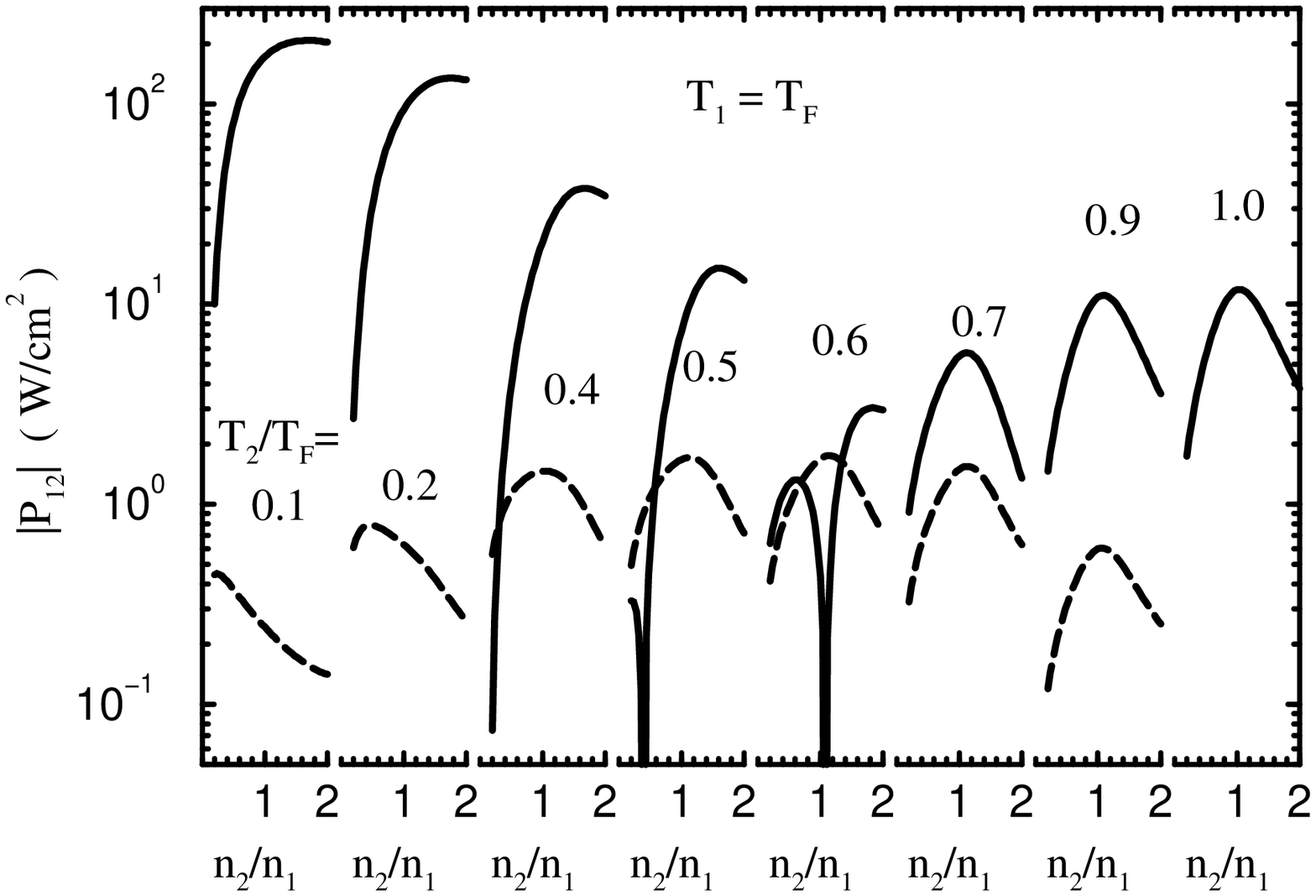}\efig{
The energy-transfer rate as functions of the ratio of
drag/drive layer electron densities $n_{2}/n_{1}$ at different
drag-layer temperatures $T_{2}$, and for electron density
$n_{1}=2\times10^{11}$\,cm$^{-2}$  $(r_{s}=1.22)$.
The dashed and solid curves are for the linear and nonlinear
($v_{1}=2E_{F}/k_{F}$) regimes, respectively. We take
$w=2a_{B}^{\star}$ and $d=5a_{B}^{\star}$. Note that in the
linear regime $P_{12}=0$ when $T_1=T_2$, regardless of the density
ratio.
}{f3}

The effect of charge density mismatch in the two layers on the 
energy-transfer rate is shown in Fig.\,3. We keep 
$n_{1}=2\times10^{11}$\,cm$^{-2}$
($r_{s}=1.22$) and $T_{1}=T_{F}$ constant, and vary $n_{2}/n_{1}$ 
in the range $0.2$ to $2$ for different $T_{2}$ values.
Both in the linear and nonlinear regimes it is observed that
$P_{12}$ is very sensitive to the electron density ratio of the 
layers. For relatively large temperatures ($T_{2}\sim T_{F}$) the 
energy-transfer is considerably larger when the densities are equal. 
In the linear regime (cf, dashed curves in Fig.\,3), the peak value 
of $P_{12}$ shifts to smaller values of $n_{2}/n_{1}$ ratio when the
temperature of the second layer is lowered, for example at 
$T_{2}=0.1\,T_{F}$ the largest transfer rates occur for 
$n_{2}\approx 0.2\,n_{1}$.
For finite drift velocities the nonlinear effects show up also
in terms of the density ratio. The vanishing of power transfer
at finite temperature differences of the layers is seen at
even lower $T_2$ if $n_2<n_1$. In the figure, for 
$v_{1}k_{F}/E_{F}=2$, the direction of energy flow is from layer 
$1$ to layer $2$ ($P_{12}<0$) if $T_{2}\le0.4T_F$, and vice versa if 
$T_{2}\ge0.7T_F$ in the chosen range of the density ratio. 
For the cases $T_2=0.5T_F$ or $0.6T_F$ however, $P_{12}<0$ at larger
values of $n_2$, and changes sign at certain values of the density
ratio, as $n_2$ is lowered. The feature is due to that 
a constant temperature correspods to a larger effective temperature
for a lower-density electron gas, in units of the Fermi temperature 
of the second layer.
The momentum
transfer rate in the drag phenomenon has been known to be very 
sensitive to the relative densities in the spatially separated 
electron systems.\cite{hill97,noh,wang} 
We find here that energy transfer rate also
has a strong dependence on the ratio $n_2/n_1$.

\bfig{h}\ff{0.45}{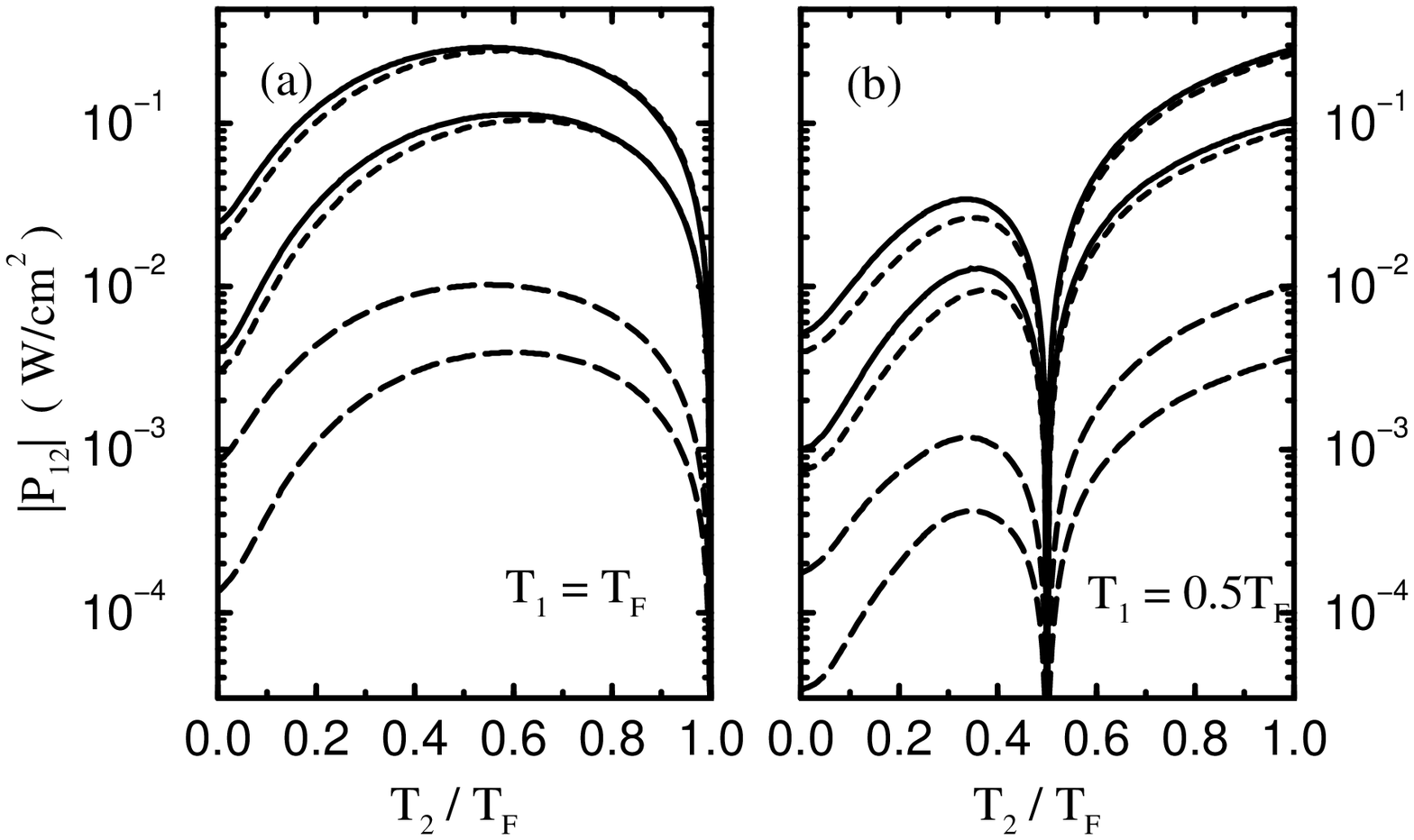}\efig{
Comparison of Coulomb and optical phonon mediated Coulomb
interaction mechanisms in the
energy transfer rate. The short- and long-dashed curves are,
respectively, for Coulomb-only and phonon-only contributions to the
energy transfer rates.
The solid curves represent the simultaneous inclusion of both
mechanisms. The upper and lower curves of each type are for
$d=5a_{B}^{\star}$ and $d=7a_{B}^{\star}$ respectively. ($n_1=n_2$,
$r_{s}=2$ and $w=2a_{B}^{\star}$)
}{f4}

Finally, we calculate the effect of optical phonons on the
energy transfer rate in a double-layer electron system.
At low temperatures, the acoustic phonons are known to
contribute to the momentum transfer rate.\cite{noh99} Since we 
are more interested in the high temperature regime where plasmons 
play a dominant role in the momentum and energy transfer rates, we
consider only the dispersionless optical phonons for the phonon
mediated interlayer electron-electron interactions. More 
specifically, we replace the interlayer
effective electron-electron interaction by
\begin{equation}
W_{12}(q,\omega)=\frac{V_{12}(q)+D_{12}(q,\omega)}
                      {\varepsilon_{T}(q,\omega)}
\end{equation}
to account for the Coulomb and LO-phonon mediated interactions 
simultaneously. The phonon term $D_{12}$ which also has a Coulombic
nature is calculated to be\cite{guven,hu98}
\begin{equation}
D_{12}(q,\omega)=V_{12}(q)
   \left(1-\frac{\epsilon_{\infty}}{\epsilon_{0}} \right)
\frac{\omega_{\rm LO}^{2}}{\omega^{2}-\omega_{\rm LO}^{2}+
i\gamma\omega}~.
\end{equation}
The above form of the interaction due to the exchange of virtual 
phonons arise when the ``bubble" diagrams to all orders are 
considered within the RPA.
Here $\gamma^{-1}$ is the phenomenological lifetime for phonons. 
$\gamma$ is taken finite in our calculations but the results do not 
have a significant dependence on $\gamma$ as long as it is small
(i.e. for GaAs, $\gamma\sim 0.1$\,meV, whereas $\omega_{\rm
LO}=36$\,meV). 
The total screening function is modified in a straightforward
way to include the phonon contribution,
\begin{eqnarray}
\varepsilon_{T}(q,\omega)&=&[1-(V_{11}+D_{11})\chi_{1}]~
                         [1-(V_{22}+D_{22})\chi_{2}] \nonumber \\
         &-& (V_{12}+D_{12})^{2}\chi_{1}\chi_{2}~.
\end{eqnarray}
$D_{ii}$ are the phonon mediated intralayer electron-electron 
interaction terms defined similar to $D_{12}$. The zeros of
the total screening function $\varepsilon_T(q,\omega)$ give
the coupled plasmon-phonon mode dispersions.
The purely Coulomb and phonon mediated interaction contributions 
to the amount of transfered energy are not simply additive, 
due to the interference terms coming from the total interaction 
amplitude, Eq.\,(7). It is possible to consider the phonon 
contribution alone by keeping only the $D_{12}$ term in the 
numerator of Eq.\,(7), i.e. 
$W_{12}(q,\omega)=D_{12}/\varepsilon_{T}$.
However one should retain\cite{bonsager98} the coupling of Coulomb 
and phonon terms in $\varepsilon_{T}(q,\omega)$.  

We calculate the energy transfer rate between two quantum
wells by modifying the expression for $P_{12}$ as set out
above for the phonon mediated Coulomb interaction. We assume
that electrons and phonons are in thermal equilibrium at
$T_1$.
In Fig.\,4 we display the energy-transfer rates due to the 
Coulomb and phonon coupling interactions alone, and when both 
mechanisms are present together. It is observed that the phonon 
contribution, being almost an order of magnitude smaller, has 
qualitatively similar form as the direct
Coulomb interaction contribution. The mentioned interference
is observable when the temperature difference of the layers is small;
the combined effect of two mechanisms can lead to slightly smaller
transfer rates (cf, Fig.\,4(a)).  

In summary, we have considered the energy transfer rate in a
double-quantum-well system in a drag experiment type setup. The
interlayer Coulomb scattering mechanism through dynamical
screening effects greatly enhances the energy transfer rate 
from one layer to another. 
We have found that a large drift
velocity corresponding to large externally applied field
greatly modifies the energy transfer rate and may lead to
power absorption from the cooler electron gas to the other.
Some of our predictions may be tested in hot-electron
photoluminescence experiments, in which the power loss of an
electron gas is measured. Lastly, we have also considered the
contribution of optical phonons in the phonon mediated
Coulomb drag which may be important to
understand the future experiments. It would be interesting 
to extend the measurements of Noh {\it et al}.\cite{noh99}
to higher temperatures to observe the effects of coupled
plasmon-phonon modes.

\acknowledgements{
This work was partially supported by the
Scientific and Technical Research Council of Turkey (T\"{U}B\.{I}TAK)
under Grant No. TBAG-2005, by NATO under Grant No. SfP971970.,
and by the Turkish Department of Defense under Grant No.
KOBRA-001. We thank Dr. P.\,J. Price for his useful comments.}

\end{document}